# Hypersymmetry of gravitational and inertial masses in relativistic field theories[1]

**György Darvas**

Symmetrion, Budapest, Hungary

*E-mail:* symmetry@symmetry.hu
orcid.org/0000-0002-1652-4138

The paper discusses, first, distinctions between gravitational and inertial masses (considered as isotopic field-charge – IFC – siblings of the gravitational field), and the ways, how they modify physical theories, including GTR. It shows that their equivalence does not mean identity. Introduction of qualitatively different mass terms modifies (among others) the gravitational equation. That leads to apparently losing its symmetry. In order to keep the symmetry of the stress-energy tensor, then, the paper identifies a symmetry group by the help of the tau algebra, which is isomorphic with the SU(2) group. The tau algebra transforms 3+1 type quantities. Invariance under the transformations of this group is called hypersymmetry (HySy). The group of HySy can make a correspondence between vector components and scalars. Next, there is shown how does the HySy group restore the apparently distorted symmetry of the stress-energy tensor. Finally, HySy is applied to the gravitational theory, and there are discussed a few consequences for the gravitational equation in quantum gravity. They are based on the, earlier disclosed, conservation of a property of the IFC-s, called isotopic field-charge spin (IFCS). It is shown that the solutions of the equations should be doubled, and another mediating boson, called dion, is to be assumed in the gravitational interaction, (anti)parallel with the graviton.

**Keywords**: hypersymmetry, relativistic field theory, conservation, (of) isotopic field charges, (in) gravitation

**PACS**: 11.30.Ly, 04.50.Kd, 04.62.+v, 04.90.+e, 02.20.Sv, 11.15.-q, 11.10.Nx

## 1 Introduction

*At first*, the paper discusses, how the distinction between *gravitational* and *inertial masses* modifies physical equations. Gravitational and inertial masses are – at least near to rest – *quantitatively* equivalent. At the same time, they are *qualitatively* different physical properties. This difference has not been reflected in the traditional physical equations. It makes itself apparent at high velocities relative to the observer, where the two kinds of masses differ also in their quantities. These two kinds of masses are considered as *isotopic field-charges* (IFC) of the gravitational interaction field (Darvas, 2011). The two, isotopic mass siblings are considered the field-sources of the energy elements in the stress-energy tensor of the gravitational field, and the kinetic (momentum like) tensor elements, respectively. Taking into account this issue, this distinction between the mass siblings involved changes also in our general view on the physical world.

Putting isotopic masses in equations destroys their Lorentz invariance, and the symmetry of the $T_{\mu\nu}$ tensor. This would contradict to our traditional precondition demanded for the physical theories, namely, a long lasting paradigm that equations of the physical interactions must be subject to invariance under the Lorentz-transformation (alone); moreover that it is not only a necessary, but also a sufficient condition. This is not the case: no physical principle stated that Lorentz-invariance is a sufficient condition demanded for the physical equations. There are other invariances that may appear together and combined with Lorentz's. However, we would like to keep the symmetry of the stress-energy tensor.

*At second*, in order to *restore the* apparently lost *invariance*, the IFC theory disclosed an additional invariance (a symmetry isomorphic with the SU(2) group). There was shown (Darvas, 2011) that IFC are subject to a hypersymmetry (HySy, i.e., conservation of a property called the *isotopic field-charge spin*, IFCS) that can transform them into each other (i.e., to rotate the IFC in an abstract IFCS field, where they can occupy two positions). It has been shown (Darvas, 2016, 2018) that HySy can be characterised by the help of an algebra (called *τ-algebra*). That symmetry guarantees to keep the covariance of our physical field equations and restores the distorted symmetry of the stress-energy tensor. HySy is broken at lower velocities (lower kinetic energies). Therefore, at least near to rest, one can observe the two IFC of the gravitational field equivalent. This indistinguishability is formulated as an equivalence principle. However, equivalence – observed among limited conditions – does not mean identity (Darvas, 2011, 2013b). We put non-identical masses in physical equations, marking the gravitational and inertial masses by different notations. Since the gravitational mass appears in the potential (scalar) part of a Hamiltonian ($V$), we denote it by $m_{gravity} = m_V$, while the inertial mass is the source of the kinetic (vector) part of a Hamiltonian ($T$), we denote it by $m_{inertia} = m_T$. So, we modified the equations (incl. the gravitational) that led to novel conclusions at high velocities, while it did not result changes near to rest. Our investigations include, how the two isotopic masses behave at high kinetic energies.

---
[1] Paper presented at the 20th International Conference "Physical Interpretations of Relativity Theory" (PIRT-2017) held at the Bauman Moscow State Technical University, 3-6 July 2017; http://www.pirt.info/?lang=eng.

*At third*, this physical description (in contrast to string theories) involves a new approach (among other interactions) to gravity. *Modification of the Gravitational Equation*[2] by inserting the two kinds of mass siblings in the stress-energy tensor, results not only in destroying its Lorentz invariance, but also in a few modifications in its solutions. Without going deep in mathematical details, the paper directs shortly the attention to a few consequences and shows, how the symmetry and invariance of the stress-energy tensor can be restored by the hypersymmetry.

In my paper presented at PIRT-2011 (Darvas, 2012a), I showed *analytically* a way, how the asymmetry hidden inside the so modified $T'_{\mu\nu}$ can be restored, along with demonstrating the conservation of the IFCS for the isotopic gravitational field charges. The same approach was successfully applied for QED in two consecutive papers (Darvas, 2013a, 2014). Now, I present an *algebraic* way that restores the invariance of the modified stress-energy tensor under an extended gauge transformation (of the IFCS) by the tau algebra and the group composed of its elements, explained in (Darvas, 2016, 2018).

**2 Isotopic masses: Isotopic charges of the gravitational field**

Distinction between the two kinds of masses, $m_{gravity} = m_V$ and $m_{inertia} = m_T$ is not new in the history of physics. Their difference was known since Newton. However, due to their quantitative equivalence (at least at rest) there was not necessary to distinct them qualitatively in the equations of physics. The experimental proof of their equivalence (first by L. Eötvös) and the formulation of the equivalence principle (by A. Einstein) inspired many textbook writers to identify them. This was pragmatically correct, but theoretically not justified. The problem became acute when technological development made possible to observe experimentally high energy interactions.

There are different approaches to handle the problem of the two kinds of masses. Let's mention first the original approach by Einstein. In the nineteen tens, he considered that according to GTR, there were only free falling bodies and inertial forces in the gravitational field, therefore the notion of the gravitational mass could be cancelled from the vocabulary of physics, and anyway, if they both existed, they were equivalent. The reason of this approach was that the left side of the gravitational equations includes only geometric properties of the curved space-time, so there is no place for masses (field sources) on the right side as well. Later he changed his mind, and noted that the individual elements of the stress-energy tensor are associated with energy and momentum (or their densities), that means, with $m_V$ and $m_T$, respectively; and the gravitational constant on the right side compensates the mass dimensions.

From our point of view, the most important approaches are that deal with the transformation of the two kinds of masses. There is an opinion that the masses do not transform with the velocity (Hraskó, 2001). The change of their inertia is due to the Lorentz transformation of their velocity, and it cannot be taken into consideration twice. Moreover, the inertia of a moving body transforms in a different proportion in the direction of its velocity (longitudinal inertial mass) and perpendicular to it (transversal inertial mass), what is an argument for leaving the Lorentz transformation solely to the velocity, and against the transformation of the mass itself, i.e., to calculate always with the rest mass. Nevertheless, this argument does not eliminate the qualitative difference between the masses $m_V$ and $m_T$.

---

[2] The *Int. J. Mod. Phys. D* published a series of papers between November 2012 and January 2014 on Modified Gravity Theories. Most of them concentrates on the left side of the Einstein equations ($F(R)$ gravity theories). A good review is given in (Myrzakulov et al., 2013).

Another approach is that of S. Fedosin (Perm), who argues repeatedly for a fixed 4/3 proportion between the masses $m_T$ and $m_V$. This approach is not acceptable, at least by the author.

A most recent another approach to make distinction between forms of masses is discussed in (Calmet and Kuntz, 2017): According to them a few observed phenomena „suggest that there is a new form of matter that does not shine in the electromagnetic spectrum. Dark matter is not accounted for by either general relativity or the standard model of particle physics. While a large fraction of the high energy community is convinced that dark matter should be described by yet undiscovered new particles, it remains an open question whether this phenomenon requires a modification of the standard model or of general relativity. Here we want to raise a slightly different question namely whether the distinction between modified gravity or new particles is always clear." They showed that this is not always the case. My comment is here to refer to the mass of the dion, mentioned in Section 6.

At the PIRT-2004, E.P.J. de Haas referred (de Haas, 2004a, 2005) to an early assumption by G. Mie (1912a, 1912b, 1913) concerning the transformation of the masses. According to Mie, the inertial mass transforms as $m_T = \kappa m_0$, and the gravitational mass transforms as $m_V = (1/\kappa)m_0$ where $m_0$ denotes the rest mass, $\kappa$ represents its Lorentz transformation. Based on these transformations, Mie formulated a version of the weak equivalence principle (never accepted by Einstein in Mie's form). As de Haas explicated in another paper (de Haas, 2004b), Mie's idea was in compliance with a later hypothesis by de Broglie (1923, 1925) on waves connected to particles with material mass. de Broglie's theory assigned wave frequencies to the moving massive particles that he called inertial clock frequencies ($v_T$) and inner-clock frequencies ($v_V$). They transform as $v_T = \kappa v_0$ and $v_V = (1/\kappa) v_0$, respectively. Although this assumption included a contradiction, de Broglie solved it by his so called "Harmony of the Phases". In that theory de Broglie first stated that the gravitational mass and the inertial mass of the same particle may possess particle- and wave-properties, respectively. (This holds even in contemporary theories.) de Haas showed that accepting the transformations assumed by Mie and de Broglie lead to the observations that, seen from a rest frame, "the equivalence of the masses is not a Lorentz-invariant condition and cannot be transformed into a fundamental axiom or law of nature"; and the same observer may conclude that "the equivalence of the phases is a Lorentz-invariant condition and that this equivalence can be seen as a fundamental law of nature". In short, he concluded that the equivalence principle of masses should be replaced by the equivalence principle of wave phases. de Haas mentioned also that answering the questions raised by the mentioned controversies is left for quantum gravity.

Accepting the usual form of the GTR (G. 't Hooft, 2002), maintaining the validity of the weak equivalence principle for masses, requires to reject the assumed Mie–de Broglie transformation rules. One can check it in an easy thought experiment. Let's imagine two, distant, different mass celestial bodies moving around their mutual mass centre at high velocities, in respect both to each other and to the Earth. Compare, how do they observe the motion of each other from their own reference frames, and how do we observe their motion from the Earth. (In fact, we can observe the period of the smaller one from the Earth.) Assuming the Mie–de Broglie transformation, we get in contradiction with the values expected by the STR and the Newton gravitational law. Our observations will coincide with those laws assuming the following transformations: $m_V = m_0$ and $m_T = \kappa m_0$, where $m_0$ is the measure of the rest mass.

According to the IFC theory, the gravitational mass does not change with velocity boost, while the inertial mass Lorentz transforms. Let's remember also a quantum gravity consideration that in the IFC theory (Darvas, 2011) interacting masses must be always in opposite IFC states: $m_V$ can interact with $m_T$ and vice versa. According to the same theory, they can be transformed into each other, since they are subject of an invariance under the transformations of a symmetry group (HySy), resulted in the IFCS conservation [analytically (Darvas, 2011), algebraically (Darvas, 2016, 2018)]. Their roles are changed permanently, like fermions do with their boson exchange. A detailed sketch of the theory was described in my PIRT-2011 paper (Darvas, 2012a), specification to the gravity in Sections 1.3 and 2.5. A Lagrangian description is given in (Darvas, 2012b, Appendix).

### 3 Isotopic masses and the stress-energy tensor

Before applying the mass siblings for the stress-energy tensor, we should make preliminary remarks. The IFC theory is relevant among strongly relativistic conditions. The derivation of the GTR included a few approximations assuming relatively weak gravitational field. Similar approximations were assumed in the solutions of the Einstein equations. And yet, seemingly, the theory works well among wide limits. One can apply the IFC theory by accepting that the Einstein equations hold (at least approximately) in the presence of strong gravitation and masses moving at high speed. Nevertheless, we should consider certain extensions of the theory to strong gravitation and the presence of a velocity dependent field effective at high energies.

The components of the stress-energy tensor can be clustered in four parts.

$$\begin{array}{c} \text{stress density} \\ \begin{bmatrix} T_{11} & T_{12} & T_{13} & | & T_{14} \\ T_{21} & T_{22} & T_{23} & | & T_{24} \\ T_{31} & T_{32} & T_{33} & | & T_{34} \\ \hline T_{41} & T_{42} & T_{43} & | & T_{44} \end{bmatrix} \Big\} \text{momentum density} \\ \underbrace{\phantom{T_{41} \; T_{42} \; T_{43}}}_{\text{energy flux density}} \quad \text{energy density} \end{array} \qquad (3.1)$$

The stress-energy tensor is symmetric. This means, the respective elements of the momentum density and the energy flux density are equal: $T_{i4} = \delta^{ik} T_{4k}$.

The assumption of the IFC theory about the difference of the gravitational and inertial masses, apparently, distorts this symmetry. There appear inertial masses ($m_T$) in the stress- and the momentum density, while there appear gravitational masses ($m_V$) in the energy flux density and the energy density. (See also a few additional remarks on this categorisation in Section 6.) The former transform with velocity, while the latter do not.

We intend to keep the symmetry of the stress-energy tensor. In this order, we should demand that under a velocity boost, the tensor transformed so that along with the invariance under the Lorentz transformation it transformed also in an invariant way under the isotopic field-charge spin (IFCS) transformation, which transforms the two kinds of masses into each other. Under field-charges, we mean here the charges of the gravitational field, $m_V$ and $m_T$. This latter transformation has been presented in several former publications by the author (Darvas, 2011; 2012a; 2012b; 2012c; 2016) and applied successfully in QED. This combined transformation guarantees to keep the symmetry of the tensor out of rest.

### 4 The group of the isotopic field-charge spin (IFCS) transformation

*4.1 The tau algebra*

There are several 3+1 parameter quantities (like vector + scalar potentials, four-currents, space-time, four-momentum, …). In most cases, the three- and the one-parameter characterised elements of these quantities differ in the field-sources (e.g., inertial and gravitational masses, Lorentz- and Coulomb-type electric charges, …) associated with them. The field-source pairs are subjects of an invariance group that can transform them into each other. We present the algebra ($\tau$) of that transformation and characterise the group of the invariance. The invariance group of the tau algebra defines a hypersymmetry (HySy).

We demonstrate matrices ($\tau$) that can function as operators in a quantum theory (that can be applied also for gravity). These operators effect state functions (in quantum gravity functionals). One can define eigenvalue equations in which the effect of the $\tau$ operators is multiplication by numbers (eigenvalues of the $\tau$ operator). The state functions in these eigenvalue equations take the form of eigenfunctions of the $\tau$ operators. Since we represent the linear operator $\tau$ in form of square matrices, we are looking for the eigenfunctions in the form of eigenvectors. We apply $\tau$ operators that are able to transform vierbeins characterised by 3+1 parameters into each other[3]. Operators $\tau$ should fulfil eigenvalue equations $\tau\varphi=k\varphi$, where $\varphi$ are eigenfunctions of the operator $\tau$, and $k$ are numbers, called eigenvalues of this equation. The tau algebra is unitary (see below), we fixed the egienvalues of $\tau$ by $\pm\frac{1}{2}$. We denoted the $\tau$ operator's eigenfunctions belonging to the eigenvalues of $\tau$ by $\varphi_+^{(\tau)}$ and $\varphi_-^{(\tau)}$, respectively. For simplicity, we denote the latter two vierbeins by $\chi$ and $\vartheta$, respectively. Having known the eigenvalues and the eigenfunctions, our task was to find operators $\tau$ that fulfil the above eigenequation.

Let's choose two eigenvectors $\chi$ and $\vartheta$ [with (+ + + -) signature] in the following representation:

$$\chi = \begin{bmatrix} 1 \\ 1 \\ 1 \\ 0 \end{bmatrix}, \quad \vartheta = \begin{bmatrix} 0 \\ 0 \\ 0 \\ i \end{bmatrix} \tag{4.1}$$

Note, we had certain limited freedom to choose the values in the eigenvectors. This choice makes handling the $\tau$ transformation-matrices convenient. Recall, that in the following case, like in many other cases, the representation of the transformation group coincides with a representation of the respective Lie algebra.

We aimed at finding matrices, which transform linearly these two eigenvectors that satisfy the eigenvalue equations for the operator $\tau$ as described below. Similar to the $\sigma$ algebra in Dirac's (1928) QED, we are looking for matrices $\tau$ with the following properties:

$$\tau_3\chi = \chi \qquad \tau_3\vartheta = -\vartheta$$
$$\tau_2\chi = \vartheta \qquad \tau_2\vartheta = \chi$$
$$\text{requiring } \tau_i\tau_j = i\tau_k \text{ } (i, j, k \text{ cyclic indices})$$
$$\tau_1\chi = -i\vartheta \qquad \tau_1\vartheta = i\chi$$

A set of the following matrices meets the required properties:

$$\tau_1 = \begin{bmatrix} 0 & 0 & 0 & 1 \\ 0 & 0 & 0 & 1 \\ 0 & 0 & 0 & 1 \\ 1 & 0 & 0 & 0 \end{bmatrix}, \quad \tau_2 = \begin{bmatrix} 0 & 0 & 0 & -i \\ 0 & 0 & 0 & -i \\ 0 & 0 & 0 & -i \\ i & 0 & 0 & 0 \end{bmatrix}, \quad \tau_3 = \begin{bmatrix} 1 & 0 & 0 & 0 \\ 1 & 0 & 0 & 0 \\ 1 & 0 & 0 & 0 \\ 0 & 0 & 0 & -1 \end{bmatrix} \tag{4.2}$$

These $\tau$ matrices satisfy the following algebra:

---

[3] In contrast to the Dirac bispinor consisting of 2+2 components (two-spinors), our spinors have a peculiarity that must be reflected in their construction. Since they are expected to make (in classical terms unusual) correspondence between *vector components* and *scalars* (c.f., also Darvas, 2016), they must be constructed in a 3+1 form. Four-columns of these bispinors can be subjects of linear transformation by [4x4] matrices whose minors should reflect that 3+1 structure. This predicts the structure and explains the character of the $\tau$-matrices.

$$\tau_1^2 = \tau_2^2 = \tau_3^2 = \begin{bmatrix} 1 & 0 & 0 & 0 \\ 1 & 0 & 0 & 0 \\ 1 & 0 & 0 & 0 \\ 0 & 0 & 0 & 1 \end{bmatrix} = \mathbf{E}, \quad \{\tau_i, \tau_j\} = 0, \quad [\tau_i, \tau_j] = 2i\tau_k \qquad (4.3)$$

Introducing the notation: $\begin{bmatrix} 1 & 0 & 0 \\ 1 & 0 & 0 \\ 1 & 0 & 0 \end{bmatrix} = \mathbf{I}_L$, $\begin{bmatrix} 0 & 0 & 1 \\ 0 & 0 & 1 \\ 0 & 0 & 1 \end{bmatrix} = \mathbf{I}_R$ the above matrices can be written in the form:

$$\tau_1 = \begin{bmatrix} 0 & \mathbf{I}_R \\ 1 & 0 \end{bmatrix}, \quad \tau_2 = \begin{bmatrix} 0 & -i\mathbf{I}_R \\ i & 0 \end{bmatrix}, \quad \tau_3 = \begin{bmatrix} \mathbf{I}_L & 0 \\ 0 & -1 \end{bmatrix} \quad \text{where} \quad \tau_1^2 = \tau_2^2 = \tau_3^2 = \mathbf{E} = \begin{bmatrix} \mathbf{I}_L & 0 \\ 0 & 1 \end{bmatrix}. \quad (4.4)$$

We denote the identity matrix of the 4 x 4 matrices by [**1**]. The unit matrix of the $\tau$ matrix-algebra is denoted by **E**.

*4.2 Group properties of the τ-matrices*

The $\tau$ matrices form a group, since they satisfy the group axioms. There are 2 independent among them. Their peculiar property is that all $\tau_i$ are the inverse group elements of themselves, that means $\tau_i^{-1} = \tau_i$ (see also 4.3 (d)). The identity element of the group is **E**.

This is the group of HySy. Representation of the transformation-group is determined by the algebra of the $\tau$-matrices.

The expressions of the $\tau$ matrices in (4.4) show very similar form to the [2 x 2] Pauli matrices:

$$\sigma_1 = \begin{bmatrix} 0 & 1 \\ 1 & 0 \end{bmatrix}, \quad \sigma_2 = \begin{bmatrix} 0 & -i \\ i & 0 \end{bmatrix}, \quad \sigma_3 = \begin{bmatrix} 1 & 0 \\ 0 & -1 \end{bmatrix}. \qquad (4.5)$$

Over the analogy, there are differences as well: in contrast to the diagonal $\sigma$ matrices, our [4x4] $\tau$-matrices manifest an apparent asymmetric character. Observe, that all the above matrices of the $\tau$-algebra are singular. In matrix algebraic terms, in general, singular matrices have neither inverse, nor adjoint matrices! This limits their commutation in scalar multiplication with other matrices. Nevertheless, in group theoretical sense, they have inverse and unit elements, considering the expressions in (4). Inverse elements of the singular matrices constituting the group are interpreted in Penrose-Moore sense and are called pseudo-inverses.

We demonstrate this on the example of the [2 x 2] $\tau$ matrices introduced in (4.4). These matrices differ from the [2 x 2] Pauli matrices that the numbers *1* in the upper row are replaced by the minor matrices $\mathbf{I}_R$ and $\mathbf{I}_L$. *In matrix algebraic terms* we should take into account the singularity of these minor matrices (and extend it to [4 x 4] matrices). *In group theoretical terms* there is not required that the constituents in the [2 x 2] matrices, forming elements of a group, should appear only as numbers. Therefore, investigating the group properties, we will not consider the special properties of $\mathbf{I}_R$ and $\mathbf{I}_L$.

*4.3 Properties of the τ matrices.*

One can check a few properties of the $\tau$ matrices by easy calculation.
(a) The $\tau$ matrices are selfadjoint, $\text{Adj}\tau_i = -\tau_i$ ($i = 1, 2, 3$).
(b) The determinants of the $\tau$ matrices coincide with those of Pauli's $\sigma$ matrices.
$\quad \tau_i \cdot \text{Adj}\tau_i = -\tau_i^2 = -\mathbf{E}$ (here *i* is not a summing index),
and comparing this with the definition, one can read from here that
$\quad \det \tau_i = -1$ ($i = 1, 2, 3$).
(c) The $\tau$ matrices are unitary. Similar to the Pauli matrices ($\sigma_i^+\sigma_i = \sigma_i\sigma_i^+ = [\mathbf{1}]$) the 2x2 tau matrices fulfil the (Hermitic sense) unitarity condition: $\tau_i^+\tau_i = \tau_i\tau_i^+ = \mathbf{E}$, where **E** represents the unit matrix of the $\tau$ algebra, in terms defined in Eq. (4.4). Since, similar to $\tau_i$, the 4x4 form of **E** is singular, the group composed by the tau matrices will be called pseudo-unitary.
(d) The $\tau$ matrices are the inverses of themselves.

$$\tau_i^{-1} = \frac{\text{Adj}\,\tau_i}{\det \tau_i} = \frac{-\tau_i}{-1} = \tau_i \;.$$

*4.4 Representation of the group composed by the τ-matrices*

Similar to the Pauli matrices, let us compose a representation of the τ-algebra in the following way:

$$\mathbf{K}_+ = (\mathbf{K}_1 + i\mathbf{K}_2); \quad \mathbf{K}_- = (\mathbf{K}_1 - i\mathbf{K}_2); \quad \mathbf{K}_3 = \frac{1}{2}\tau_3$$

where $\mathbf{K}_i = \frac{1}{2}\tau_i$, so

$$\mathbf{K}_+ = \begin{bmatrix} 0 & 0 & 0 & 1 \\ 0 & 0 & 0 & 1 \\ 0 & 0 & 0 & 1 \\ 0 & 0 & 0 & 0 \end{bmatrix} = \begin{bmatrix} 0 & \mathbf{I}_R \\ 0 & 0 \end{bmatrix}; \quad \mathbf{K}_- = \begin{bmatrix} 0 & 0 & 0 & 0 \\ 0 & 0 & 0 & 0 \\ 0 & 0 & 0 & 0 \\ 1 & 0 & 0 & 0 \end{bmatrix} = \begin{bmatrix} 0 & 0 \\ 1 & 0 \end{bmatrix};$$

$$\mathbf{K}_3 = \begin{bmatrix} \tfrac{1}{2} & 0 & 0 & 0 \\ \tfrac{1}{2} & 0 & 0 & 0 \\ \tfrac{1}{2} & 0 & 0 & 0 \\ 0 & 0 & 0 & -\tfrac{1}{2} \end{bmatrix} = \begin{bmatrix} \mathbf{I}_L/2 & 0 \\ 0 & -\tfrac{1}{2} \end{bmatrix}$$

where

$$[\mathbf{K}_3, \mathbf{K}_+] = \mathbf{K}_+; \quad [\mathbf{K}_3, \mathbf{K}_-] = -\mathbf{K}_-; \quad [\mathbf{K}_+, \mathbf{K}_-] = 2\mathbf{K}_3;$$

$$\mathbf{K}_+\mathbf{K}_- = \begin{bmatrix} 1 & 0 & 0 & 0 \\ 1 & 0 & 0 & 0 \\ 1 & 0 & 0 & 0 \\ 0 & 0 & 0 & 0 \end{bmatrix} = \begin{bmatrix} \mathbf{I}_L & 0 \\ 0 & 0 \end{bmatrix}; \quad \mathbf{K}_-\mathbf{K}_+ = \begin{bmatrix} 0 & 0 & 0 & 0 \\ 0 & 0 & 0 & 0 \\ 0 & 0 & 0 & 0 \\ 0 & 0 & 0 & 1 \end{bmatrix} = \begin{bmatrix} 0 & 0 \\ 0 & 1 \end{bmatrix}$$

$$\mathbf{K}_+ + \mathbf{K}_- = \tau_1; \quad \mathbf{K}_+ - \mathbf{K}_- = i\tau_2; \quad \mathbf{K}_+\mathbf{K}_- - \mathbf{K}_-\mathbf{K}_+ = \tau_3; \quad \mathbf{K}_+\mathbf{K}_- + \mathbf{K}_-\mathbf{K}_+ = \mathbf{E}$$

(a) $\mathbf{K}^2$ is Casimir invariant. Composing the sum $\mathbf{K}^2 = \mathbf{K}_1^2 + \mathbf{K}_2^2 + \mathbf{K}_3^2 = \frac{1}{2}(\mathbf{K}_+\mathbf{K}_- + \mathbf{K}_-\mathbf{K}_+) + \mathbf{K}_3^2 = \frac{1}{4}(\tau_1^2 + \tau_2^2 + \tau_3^2) = \frac{3}{4}\mathbf{E}$ we get that $\mathbf{K}^2$ is proportional to the identity map (in our case the identity is represented by the unit matrix $\mathbf{E}$ of the group), it is a Casimir invariant that commutes with all the three generators of the representation: $[\mathbf{K}^2, \mathbf{K}_i] = 0$ ($i = 1, 2, 3$).

(b) The eigenvalues belonging to the $\mathbf{K}$ operators are half-integer. We denote the weight of this representation belonging to the eigenvector $\chi$ by $k^{(\chi)}$ and that belonging to the eigenvector $\vartheta$ by $k^{(\vartheta)}$.

$$\mathbf{K}_3 \chi = \frac{1}{2}\tau_3 \chi = \frac{1}{2}\mathbf{I}_L \chi = k^{(\chi)} \chi; \quad k^{(\chi)} = \frac{\mathbf{I}_L}{2}$$

$$\mathbf{K}_3 \vartheta = \frac{1}{2}\tau_3 \vartheta = -\frac{1}{2}\vartheta = k^{(\vartheta)} \vartheta; \quad k^{(\vartheta)} = -\frac{1}{2}$$

The effects of the $\mathbf{K}_\pm$ operators on the eigenvectors are:

$\mathbf{K}_+\chi = 0, \; \mathbf{K}_-\chi = -i\vartheta, \; \mathbf{K}_+\vartheta = \chi, \; \mathbf{K}_-\vartheta = 0.$

This representation shows similitude to a representation of the Pauli matrices, considering the same restrictions for the $\mathbf{I}_L$ and $\mathbf{I}_R$ minor matrices like above.

*4.5 Identifying the group*

All properties of this representation *formally coincide* with a representation of the SU(2) group of the Pauli $\sigma$ matrices. In a similar way, it is a special unitary group with two independent parameters, with the condition that its unitary group element is defined as $\mathbf{E} = \begin{bmatrix} \mathbf{I}_L & 0 \\ 0 & 1 \end{bmatrix}$, (whose $\mathbf{I}_L$ was given in (4.4) and is a singular minor matrix in the 4 x 4 $\mathbf{E}$). In the latter, restricted sense the group formed by the $\tau$-matrices is a special, *pseudo-unitary,* 2 parametric group. We showed that this special, pseudo-unitary group is *isomorphic* with the usual *SU(2)* group (e.g., among others of the Pauli matrices). The $\tau$-matrices behave like spinor elements of an SU(2) group.

The peculiarities of the $\tau$-matrices (that distinguish them from the Pauli matrices) allow them to make *correspondence between scalars* and *vector components*.

As shown in (Darvas, 2016, 2018) pseudo-unitary groups may take the form U(*n-l, l*). In case of the group of our $\tau$ operators, $n = 2$, $l = 0$ or 1. The corresponding groups are either U(2, 0) = U(2) or U(1,1). The latter option may arise in case of a $\tau_3$ representation (whose main diagonal contains two opposite signed elements, i.e., $l = 1$). U(1,1) needs to use an infinite-dimensional Hilbert space, because being a non-compact Lie group, U(1,1) does not admit a finite-dimensional unitary representation. Consequently, the pseudo-unitary group of the $\tau$ operators is isomorphic with the unitary U(2) group. As it was shown above, one can assign a |1| absolute value determinant to the (otherwise singular) $\tau$-matrices. In this sense, *the $\tau$ operators form a group that is isomorphic with the special unitary SU(2) group*.

### 4.6 *The $\tau$ and the Dirac ($\gamma$) algebra*

Dirac assigned a set of [4 x 4] $\rho_i$ matrices to the [4 x 4] bispinor $\sigma_i$ matrices by interchanging the second and third rows, and the second and third columns. His $\gamma$ algebra (called generally the Dirac algebra) was defined by the $\gamma_i = \rho_2 \sigma_i$ matrices. (Note, there appeared other representations of the Dirac algebra later.)

Notice that the second and third rows and columns of our $\tau$-matrices coincide. Therefore, when we assign a set of matrices to the $\tau_i$ matrices – using the analogy of Dirac's $\rho_i$ matrices – they coincide with the $\tau_i$ matrices themselves. Instead of a $\rho$-$\sigma$ pair, we've got a $\tau$-$\tau$ pair[4].

Consequently, if we define – in a similar way like we obtained the $\gamma$ matrices – a set of T [read: upper case Greek tau] matrices, we get – by definition – the following: $T_i = \tau_2 \tau_i$. Since the $\tau$-matrices transform into each other, we get the following algebra:

$$\begin{aligned} T_1 &= \tau_2 \tau_1 = -i\tau_3 \\ T_2 &= \tau_2 \tau_2 = \mathbf{E} \quad \text{(left handed „unit" operator)} \\ T_3 &= \tau_2 \tau_3 = i\tau_1 \end{aligned} \qquad (4.6)$$

Let us define $T_4 = \tau_4^2 = \tau_3$, where $\tau_4 = \begin{bmatrix} 1 & 0 & 0 & 0 \\ 1 & 0 & 0 & 0 \\ 1 & 0 & 0 & 0 \\ 0 & 0 & 0 & i \end{bmatrix}$ (4.7)

This set defined another representation of the *tau* algebra that we denote by the upper case Greek *T*. (The *T*-algebra should behave in a similar way over the abstract field in which the isotopic field-charges are rotated, like the $\gamma$ algebra did over the bispinors' space.)

Applying the above defined *T*-matrices expressed with $\tau$'s, we got the following algebra for the transformation of the eigenvectors $\chi$ and $\vartheta$:

---

[4] (1) The structure of the expected eigenvectors ($\chi$ and $\vartheta$) determine the structure of the $\tau$-matrices. (2) The analogy to the $\rho$-$\sigma$ matrix set doublet demands that the change of the second rows and columns produce unchanged $\tau$-matrices to achieve identical matrices and avoid duplication, unlike the $\rho$-$\sigma$ pair. (3) $\tau$ is the next letter in the Greek alphabet follwing $\rho$ and $\sigma$ – this justifies the name of the $\tau$-matrices. (4) The rank 2 of the matrices in the $\tau$-algebra is not surprising. These matrices make correspondence between two physical quantities. One of these quantities is a *one-component* scalar, and the other is a vector – constructed by *three-components*. This is reflected in the construction of the $\tau$-matrices.

$$T_1\chi = -i\chi \qquad T_1\vartheta = i\vartheta$$
$$T_2\chi = \chi \qquad T_2\vartheta = \vartheta$$
$$T_3\chi = \vartheta \qquad T_3\vartheta = -\chi$$
$$T_4\chi = \chi \qquad T_4\vartheta = -\vartheta$$

Let's introduce the following notations: $\delta_1 = \tau_1^2; \quad \delta_2 = \tau_2^2; \quad \delta_3 = \tau_3^2; \quad \delta_4 = \tau_4^2$.

*4.7 The algebra of the δ- and the τ-matrices*

Similar to the K representation of the $\tau$-matrices, let's compose an L representation of the $\delta$-matrices.

The effects of the $\delta_1 = \delta_2 = \delta_3 = \delta_i = \tau_i^2 = \mathbf{E}$ and the $\delta_4 = \tau_3$ matrices on the $\chi$ and $\vartheta$ eigenvectors are:

$$\delta_i \chi = \chi \quad \delta_i \vartheta = \vartheta$$
$$\delta_4 \chi = \chi \quad \delta_4 \vartheta = -\vartheta \tag{4.8}$$

Define:

$$\mathbf{L}_+ = \frac{1}{2}(\delta_i + \delta_4) = \begin{bmatrix} I_L & 0 \\ 0 & 0 \end{bmatrix} = \mathbf{K}_+\mathbf{K}_-$$

$$\mathbf{L}_- = \frac{1}{2}(\delta_i - \delta_4) = \begin{bmatrix} 0 & 0 \\ 0 & 1 \end{bmatrix} = \mathbf{K}_-\mathbf{K}_+$$

then:

$$\mathbf{L}_+ + \mathbf{L}_- = \mathbf{E} = \delta_i \quad \text{and} \quad \mathbf{L}_+ - \mathbf{L}_- = \tau_3 = \delta_4 \quad (\text{where } i = 1, 2, 3)$$

The $\mathbf{L}_+$ and $\mathbf{L}_-$ operators are represented by projector matrices $\mathbf{L}_+^2 = \mathbf{L}_+; \mathbf{L}_-^2 = \mathbf{L}_-$. Further, in contrast to the $\mathbf{K}$ operators, they are orthogonal: $\mathbf{L}_+\mathbf{L}_- = \mathbf{L}_-\mathbf{L}_+ = 0$. Note, that the $\mathbf{L}_+$ and $\mathbf{L}_-$ operators are expressed in terms of a $\tau_3$ representation of the tau algebra (by the help of its unit matrix $\mathbf{E}$ and the $\tau_3$ matrix).

The $\delta_i$ (=$\mathbf{E}$) and $\delta_4$ (=$\tau_3$) matrices compose a group, in contrast to the $\gamma$ matrices of the Dirac algebra that do not. The effects of the $\mathbf{L}_+$ and $\mathbf{L}_-$ matrices on the $\chi$ and $\vartheta$ eigenvectors are:

$$\mathbf{L}_+\chi = \chi \quad \mathbf{L}_+\vartheta = 0$$
$$\mathbf{L}_-\chi = 0 \quad \mathbf{L}_-\vartheta = \vartheta$$

that means, $\mathbf{L}_+$ acts only on $\chi$ and the orthogonal $\mathbf{L}_-$ acts only on $\vartheta$. Their effect is the opposite of the effects of the $\mathbf{K}_\pm$ operators. However, the effects of the $\delta$-matrices and the $\mathbf{L}$ matrix representation do not modify the ±1/2 eigenvalues of the $\chi$ and $\vartheta$ eigenvectors revealed afore.

The effect of the two types of $\delta$-matrices on the eigenvectors $\chi$ and $\vartheta$ of the operators of the $\tau$-algebra can be expressed in the following short [8 x 8] matrix form, where $\delta_i$ and $\delta_4$ are 4x4 matrices that act on the four-element vector columns of $\chi$ and $\vartheta$:

$$\begin{bmatrix} \delta_i & 0 \\ 0 & \delta_4 \end{bmatrix} \begin{bmatrix} \chi & \vartheta \\ \chi & \vartheta \end{bmatrix} = \begin{bmatrix} \chi & \vartheta \\ \chi & -\vartheta \end{bmatrix}$$

The $\delta_i$ matrices leave intact the eigenvectors $\chi$ and $\vartheta$, while $\delta_4$ changes the sign of one of the eigenvectors $\vartheta$.

*4.8 Conclusions*

These $\delta_i$ (i = 1, 2, 3) and $\delta_4$ (together $\delta_\nu$; $\nu$ = 1, ..., 4) matrices extend the transformation matrices in the field equations modified with the introduction of IFCs, in order to restore their covariance lost by the introduction of isotopic field-charges. The transformation matrices of the original field equations should be replaced by products of the respective $\delta$ and traditional transformation matrices.

We showed that there exists a hypersymmetric invariance group that can transform sources of 3+1 element physical quantities into each other. The group of the $\tau$-matrices can make a unique correspondence between vector components and scalars. This group is isomorphic with the special unitary SU(2) group. This group can make a correspondence between such isotopic physical quantity siblings (Darvas, 2017) like the inertial and gravitational masses, Lorentz type and Coulomb type electric charges, etc., as predicted analytically in (Darvas, 2011; 2012a; 2012b; 2012c; 2013a; 2013b; 2014). The group $\tau$ defines an invariance between particles that compose symmetric pairs. These particle twin siblings (hyperpairs) differ in their nature and physical properties from those predicted in the SUSY. Therefore, for the sake of distinction, we call their invariance HySy.

**5 Application of the tau algebra for the gravitational stress-energy tensor**

Let's return to the distortion of the stress-energy tensor $T_{\mu\nu}$ by the introduction of the isotopic masses $m_V$ and $m_T$. Let $T_{\mu\nu} = t''_{\mu\nu}$ and agree to tuck all constants into $t''_{\mu\nu}$, but not all parameters. We will write one by one those parameters and variables that are important in our aspect. Observed from a moving reference frame, the $T_{\mu\nu}$ tensor components Lorentz transform by $\kappa$; let's tuck these $\kappa$ also into $t''_{\mu\nu}$. The components $T_{i4}$ can be expressed as $t''_{i4}ic\delta^{il}p_l(m_{Tl})$; and the components $T_{4k}$ can be expressed as $t''_{4k}E_k(m_V)/c$. Further, since the individual components of the stress-energy tensor denote densities, in the following we will apply mass densities ($\rho$) instead of masses, and tuck the denominators also in the $t''_{\mu\nu}$-s. For simplicity, we hide the mass density dependence also into the $t''_{ik}$-s in the stress density sector ($i, k = 1, 2, 3$). We are interested, first of all, in the components in the momentum density ($T_{i4}$) and the energy flux density ($T_{4k}$) sectors, therefore we write these components in more detail. Their individual components can be written as $t''_{i4}ic\delta^{il}p_l(\rho_{Tl})$ and $t''_{4k}E_k(\rho_V)/c$, respectively. Note that the respective $icp_i$ and $E_k/c$ tensor components denote qualitatively different quantities, not only due to the difference of the included $m_V$ and $m_T$ masses, but also due to their transformation rules (cf. Hraskó, 2001, Section 2.8, especially his Eq. (2.8.3). Here $p_i(\rho_{Ti})$ denotes that the $p_i$ momentum density components depend on the respective inertial mass components. The index $i$ in $\rho_{Ti}$ denotes the individual mass component projections in the directions of the velocity of the moving reference frame. We must distinguish them, because the same inertial mass $m_T$ transforms by $\kappa^3$ in the direction of the velocity (longitudinal mass), and by $\kappa$ in perpendicular directions (transversal mass). For we will not treat these different transformations in the following, we will also tuck these $\kappa^3$ and $\kappa$ into $t'_{\mu\nu}$-s. We will denote them in the following: $t''_{i4}icp_i(\rho_{Ti}) = t'_{i4}icp_i(\rho_T)$, where $\rho_T$ denotes the rest value of the inertial mass density and is the same in all directions. We do not need to make similar tucking in the energy density flux, since the gravitational mass densities ($\rho_V$) do not transform with the velocity, thus we can apply the following replacement, using $E_k/c = \rho_V c$: $t''_{4k}E_k(\rho_V)/c = t'_{4k}\rho_V c$. For distinction from the mass densities, we will denote the total energy density in $t_{44}$ by the index $E$, (i.e., $\rho_E$).

Applying all these notations, the stress-energy tensor (3.1) can be now rewritten with the listed simplified notations in the following form:

$$T_{\mu\nu} = \begin{bmatrix} t''_{11} & t''_{12} & t''_{13} & t''_{14}icp_1(\rho_{T1}) \\ t''_{21} & t''_{22} & t''_{23} & t''_{24}icp_2(\rho_{T2}) \\ t''_{31} & t''_{32} & t''_{33} & t''_{34}icp_3(\rho_{T3}) \\ t''_{41}E_1(\rho_V)/c & t''_{42}E_2(\rho_V)/c & t''_{43}E_3(\rho_V)/c & t''_{44}\rho_E(\rho_V) \end{bmatrix}$$

$$T_{\mu\nu} = \begin{bmatrix} t'_{11} & t'_{12} & t'_{13} & t'_{14}icp_1(\rho_T) \\ t'_{21} & t'_{22} & t'_{23} & t'_{24}icp_2(\rho_T) \\ t'_{31} & t'_{32} & t'_{33} & t'_{34}icp_3(\rho_T) \\ t'_{41}\rho_V c & t'_{42}\rho_V c & t'_{43}\rho_V c & t'_{44}\rho_E \end{bmatrix} \quad (5.1)$$

As mentioned in Section 3, we would like to keep the symmetry of the stress-energy tensor. The expressions of the components in the momentum density and in the energy flux density can be equal if the isotopic mass terms (or their densities) appearing in them can be transformed into each other. We must demonstrate, that when transposing the tensor, the qualitatively different $T_{i4}$ tensor components can be transformed into $T_{4k}$ so that we get equal quantities. In other words, our goal is to show that there is a transformation which can transform the asymmetric $T_{i4}$ and $T_{4k}$ elements into each other.

Matrices of the group $\tau$ can provide this transformation. However, they do not transform directly the components of the stress-energy tensor, rather the eigenfunctions that satisfy quantum gravitational field equations (see Section 6). We see, the components in subject differ in the mass density terms and their velocity dependence. To show that they can be transformed into each other, first we must demonstrate that mass terms can be separated from other coefficients in the respective tensor elements.

In order to understand the relation, we illustrate this on a semi-classical example of virtual force components, expressed by the multiplication of the stress-energy tensor and a mass-density current $F^\mu = T_{\mu\nu} J^\nu$. Note, this force expresses the result of an interaction between a gravitational field with another mass' density current. The masses in the tensor and the current belong to two interacting systems, therefore, we will distinguish their mass densities with parenthetic upper indices (1) and (2). $u^k$ denote contravariant velocity vector components of the mass density current moving in respect to the gravitational field characterised by the tensor $T_{\mu\nu}$.

$$F^\mu_{virt} = T_{\mu\nu} J^\nu = \begin{bmatrix} t'_{11} & t'_{12} & t'_{13} & t'_{14} ic p_1(\rho^{(1)}_T) \\ t'_{21} & t'_{22} & t'_{23} & t'_{24} ic p_2(\rho^{(1)}_T) \\ t'_{31} & t'_{32} & t'_{33} & t'_{34} ic p_3(\rho^{(1)}_T) \\ t'_{41} \rho^{(1)}_V c & t'_{42} \rho^{(1)}_V c & t'_{43} \rho^{(1)}_V c & t'_{44} \rho_E \end{bmatrix} \begin{bmatrix} \rho^{(2)}_{T1} u^1 \\ \rho^{(2)}_{T2} u^2 \\ \rho^{(2)}_{T3} u^3 \\ ic \rho^{(2)}_V \end{bmatrix} =$$

$$= \begin{bmatrix} t'_{11} \rho^{(2)}_{T1} u^1 + t'_{12} \rho^{(2)}_{T2} u^2 + t'_{13} \rho^{(2)}_{T3} u^3 - t'_{14} c^2 p_1(\rho^{(1)}_T) \rho^{(2)}_V \\ t'_{21} \rho^{(2)}_{T1} u^1 + t'_{22} \rho^{(2)}_{T2} u^2 + t'_{23} \rho^{(2)}_{T3} u^3 - t'_{24} c^2 p_2(\rho^{(1)}_T) \rho^{(2)}_V \\ t'_{31} \rho^{(2)}_{T1} u^1 + t'_{32} \rho^{(2)}_{T2} u^2 + t'_{33} \rho^{(2)}_{T3} u^3 - t'_{34} c^2 p_3(\rho^{(1)}_T) \rho^{(2)}_V \\ t'_{41} c \rho^{(1)}_V \rho^{(2)}_{T1} u^1 + t'_{42} c \rho^{(1)}_V \rho^{(2)}_{T2} u^2 + t'_{43} c \rho^{(1)}_V \rho^{(2)}_{T3} u^3 + t'_{44} ic \rho_E \rho^{(2)}_V \end{bmatrix}$$

Now, let we introduce again a simplification: tuck the velocity dependence of the mass densities $\rho^{(2)}_{Ti}$ also in the $t'_{\mu k}$-s (k = 1, 2, 3). They look with the shortened notation: $t'_{\mu k} \rho^{(2)}_{Ti} = t_{\mu k} \rho^{(2)}_T$. Now, we can disassemble this virtual force vector to the multiplication of a [4x4] matrix and a vierbein consisting of mass densities:

$$F^\mu_{virt} = \begin{bmatrix} t_{11} u^1 & t_{12} u^2 & t_{13} u^3 & t_{14} ic^2 p^{(1)}_1 \\ t_{21} u^1 & t_{22} u^2 & t_{23} u^3 & t_{24} ic^2 p^{(1)}_2 \\ t_{31} u^1 & t_{32} u^2 & t_{33} u^3 & t_{34} ic^2 p^{(1)}_3 \\ t_{41} c \rho^{(1)}_V u^1 & t_{42} c \rho^{(1)}_V u^2 & t_{43} c \rho^{(1)}_V u^3 & t_{44} c \rho_E \end{bmatrix} \begin{bmatrix} \rho^{(2)}_T \\ \rho^{(2)}_T \\ \rho^{(2)}_T \\ i \rho^{(2)}_V \end{bmatrix} \quad (5.2)$$

The separability of the mass densities is demonstrated (at least in the case of the mass densities of the current). We saw in the Section 4.1 that $\tau_2$ can transform such vierbeins into each other:

$$\tau_2 \begin{bmatrix} \rho_T^{(2)} \\ \rho_T^{(2)} \\ \rho_T^{(2)} \\ i\rho_V^{(2)} \end{bmatrix} = \begin{bmatrix} \rho_V^{(2)} \\ \rho_V^{(2)} \\ \rho_V^{(2)} \\ i\rho_T^{(2)} \end{bmatrix} \qquad (5.3)$$

and back.

We have any reason to assume, that the mass dependence of the components of $T_{\mu\nu}$ is linear, and does not include either lower, or higher powers of $m$ than 1, namely $t'_{\mu\nu}f(m)$, where $f(m)$ is a function of the respective isotopic mass. Provided that in (5.1) all $t'_{\mu\nu}(\rho^{(1)}{}_\lambda)$ ($\lambda = T, V$) have the linear multiplication form $t'_{\mu\nu}(\rho^{(1)}{}_\lambda) = t^{\#}_{\mu\nu} \rho^{(1)}{}_\lambda$, those $\rho^{(1)}{}_\lambda$ can also be detached from $T_{\mu\nu}$, and a similar $\tau_2$ transformation applies to them.

According to the IFC theory, inertial masses interact always with gravitational ones, and vice versa. This holds in our above example for the momentum density and the energy flux density sectors. Let's now discuss the rest of the tensor. The right side of (5.2) can be written in the form:

$$\begin{bmatrix} t_{11}u^1 & t_{12}u^2 & t_{13}u^3 & t_{14}ic^2 p_1^{(1)} \\ t_{21}u^1 & t_{22}u^2 & t_{23}u^3 & t_{24}ic^2 p_2^{(1)} \\ t_{31}u^1 & t_{32}u^2 & t_{33}u^3 & t_{34}ic^2 p_3^{(1)} \\ t_{41}c\rho_V^{(1)}u^1 & t_{42}c\rho_V^{(1)}u^2 & t_{43}c\rho_V^{(1)}u^3 & t_{44}c\rho_E \end{bmatrix} \begin{bmatrix} \rho_T^{(2)} \\ \rho_T^{(2)} \\ \rho_T^{(2)} \\ i\rho_V^{(2)} \end{bmatrix} = \begin{bmatrix} [t_{ik}u^k] & t_{i4}ic^2 p_i^{(1)} \\ t_{4k}c\rho_V^{(1)}u^k & t_{44}c\rho_E \end{bmatrix} \begin{bmatrix} \rho_T^{(2)} \\ i\rho_V^{(2)} \end{bmatrix}$$

$$\begin{bmatrix} [t_{ik}u^k] & t_{i4}ic^2 p_i^{(1)} \\ t_{4k}c\rho_V^{(1)}u^k & t_{44}c\rho_E \end{bmatrix} \begin{bmatrix} \rho_T^{(2)} \\ i\rho_V^{(2)} \end{bmatrix} = \begin{bmatrix} [t_{ik}u^k] & 0 \\ 0 & t_{44}c\rho_E \end{bmatrix} \begin{bmatrix} \rho_T^{(2)} \\ i\rho_V^{(2)} \end{bmatrix} + \begin{bmatrix} 0 & t_{i4}ic^2 p_i^{(1)} \\ t_{4k}c\rho_V^{(1)}u^k & 0 \end{bmatrix} \begin{bmatrix} \rho_T^{(2)} \\ i\rho_V^{(2)} \end{bmatrix} \qquad (5.4)$$

The second component of the sum in the right side of (5.4) is in concordance with the IFC theory. The first component in the right side of (5.4) includes the stress density and the full energy density sectors. The former contains stress components that, in principle by their names, are inertial, and are here to be multiplied with the inertial part of the current's mass density vector. The full energy density that depends, in principle, on the gravitational mass is to be multiplied with the gravitational part of the current's mass density vector. However, these multiplications do not mean real interaction. They give additions to the value of the calculated virtual force, but do not represent the interaction itself. Moreover, due to shear, the stress section contains entangled contributions by the scalar and vector parts of the field strengths. The full energy density in $T_{44}$ includes the contributions of all masses constituting the field, and is an invariant, i.e., does not transform under interaction.

**6 Hypersymmetry of the gravitational equations**

The original gravitational equation is invariant under the Lorentz transformation, however, this property does not ensure sufficiency for the covariance of the equation under strongly relativistic conditions. As many publications state, the theory needs certain modifications (cf. footnote 1). Among others, (Wüthrich et al., 2012) gives a good review on approaches to quantum gravity (first of all, from the aspect of the "lost" time in some approaches to it). Covariance under the Lorentz transformation should be extended. Such an extension is discussed in this Section. (Darvas, 2011) showed that $\Delta$ (IFCS) is a two valued, spin-like property that appears in the presence of a kinetic gauge field at high energies. In order to recall: The two isotopic states are associated with the field-charges appearing in the scalar (potential, $V$) and in the vector (kinetic, $T$) parts of a Hamiltonian, respectively. Qualitative distinction was made between the two (isotopic) kinds of field-charges. A given particle can occupy either the one or the other IFC-state.

The wave function of a given particle may be in a potential state with amplitude $\psi_V$, or in a kinetic state with amplitude $\psi_T$.[5] The wave function of a single physical particle is an entanglement of the probabilities being in the potential or the kinetic state at a given moment: $\psi = \begin{pmatrix} \psi_T \\ \psi_V \end{pmatrix}$, where $\psi_T$ is a three-component column.

In quantum gravity, $\psi$ can be interpreted, e.g., as subject of a formally Schrödinger-like but in fact Wheeler-deWitt equation (WdW, with Hamiltonian constraint) that contains information about both the geometry and the matter content of the universe. It is a functional that includes space-time and configurations of other physical fields over it. It is (re)parametrised not only by the space and time parameters separated, but also by the $\Delta$ parameters.

As regards the latter ones, we denoted the eigenfunctions $\psi_T$ and $\psi_V$ that belong to the two eigenvalues of the operator $\boldsymbol{\tau}$ by $\varphi_+^{(\tau)}$ and $\varphi_-^{(\tau)}$, respectively, in Section 4.1. The operators $\boldsymbol{\tau}$ should fulfil eigenvalue equations $\boldsymbol{\tau}\varphi = k\varphi$, where $\varphi$ are eigenfunctions of the operator $\boldsymbol{\tau}$, and $k$ are numbers. There belong two isotopic field-charge spin ($\Delta$) positions to the sources of the two states, in accordance with previous papers (Darvas, 2011, 2016). During transition from $\psi_V$ to $\psi_T$ and back, the source of the field, i.e., the respective field-charge, needs to change its $\Delta$ state. The operator $\boldsymbol{\tau}$, whose matrix algebra was introduced in the Section 4 affects those $\Delta$ states.

First, recall that $\chi$ and $\vartheta$ are the eigenfunctions of the $\Delta$ isotopic field-charge spin's $\boldsymbol{\tau}$ operators. (As mentioned, the corresponding eigenfunctions were denoted also by $\varphi_+^{(\tau)}$ and $\varphi_-^{(\tau)}$). The eigenvalue equations are $\boldsymbol{\tau}\varphi_+^{(\tau)} = \frac{1}{2}\varphi_+^{(\tau)}$ or $\boldsymbol{\tau}\chi = \frac{1}{2}\chi$ and $\boldsymbol{\tau}\varphi_-^{(\tau)} = -\frac{1}{2}\varphi_-^{(\tau)}$ or $\boldsymbol{\tau}\vartheta = -\frac{1}{2}\vartheta$, respectively. The eigenvalues of $\Delta$ can take $\pm\frac{1}{2}$. The operator $\boldsymbol{\tau}$ refers to the transformation matrices of the HySy *group* discussed in Sec. 4 that rotate, e.g., the IFCS in a velocity dependent kinetic gauge field. Since the operators responsible for the rotation in the kinetic gauge field do not affect the space and time dependent (and other dependence) components in the field equations, and vice versa, the particle's full state function $\psi$ can be separated according to the following sum: $\psi = \psi'\varphi_+^{(\tau)} + \psi''\varphi_-^{(\tau)}$, where $\psi'$ and $\psi''$ denote the state functions affected by the space-time dependent operators. The $\psi$ state functions in (6.1) below are functions of the $x^\nu$ space-time coordinates, as well as of the $\Delta = \pm\frac{1}{2}$ parameters (according to the set of eigenvalues of the $\boldsymbol{\tau}$ operators). The differential operators act only on $\psi(x^\nu)$, and the $\boldsymbol{\tau}$ operators on the eigenfunctions of $\Delta$. This is a useful property, because the doubled solutions can be obtained by extending the space-time dependent state functions with IFCS ($\Delta$) dependent functions as multipliers.

Thus, the $\psi = \psi(x^\nu, \Delta)$ state function can be dissociated to the following products according to the two eigenfunctions of the $\boldsymbol{\tau}$ matrix operators:

$$\psi(x^\nu, \Delta) = \psi'(x^\nu)\varphi_+^{(\tau)}(\Delta) + \psi''(x^\nu)\varphi_-^{(\tau)}(\Delta) \tag{6.1}$$

This involves also that the $\psi$ state functions in each of the original Einstein equations' solutions can be separated into two further space-time dependent functions, in accordance with the additional bivariant opposite positions of the IFCS. (Important to notice: the coincidence of the set of matrices in the $\tau$- and the $T$-algebras ensures that the number of solutions is *doubled* thanks to the IFCS invariance, according to the two new degrees of freedom brought in by the two possible positions of the IFCS – compared with the *quadruple* solutions of the Dirac equation due to the new degrees of freedom by the **ρ** and **σ** operators (Dirac, 1928, 1929).) It can be exemplified so that, according to the two isotopic states of the field-charges, two separated solutions are assigned to each of the known solutions of the gravitational equations. The $\psi$ functions satisfy $\hat{H}\psi(x^\nu, \Delta) = 0$ where $\hat{H}$ is a WdW Hamiltonian (and in which form the time dependence of $\psi$ disappears). According to the IFCS theory, the graviton exchange, which is a result of the symmetry of the space-time part of the state function, must be accompanied by the exchange of another boson, called dion, due to the symmetry of the $\varphi^\tau$ part of the state function(al) under the transformations of the **τ** group, which is isomorphic with the SU(2) group. In short, they are invariant under HySy transformations. The dion must carry the mass difference between the gravitational and the Lorentz transformed inertial masses (Darvas, 2011). We mention also that while the operators **τ** may rotate the masses (mass densities) among themselves, they leave the rest of the state functions intact; moreover they guarantee to save the real values in the stress-energy tensor, and the negativity of the $T_{44}$ component.

---

[5] According to one of the interpretations of the IFCS theory, the energy of a single particle at a given moment is considered to be ideally concentrated either in the scalar or in the kinetic part of its Hamiltonian (Darvas, 2011). The corresponding state functions belonging to these two idealised reduced Hamiltonians are denoted by $\psi_V$ and $\psi_T$. They characterise the s.c. *potential* and *kinetic* states of the particle. The particle takes one of these two states at certain *probabilities*. According to this interpretation the particle can change its state (or oscillate) between the fully potential and the fully kinetic states. The observable state function can be characterised by the mixture of these two probability states and denoted by $\psi$. The IFCS theory describes transition between the two states.

Recall that the transformation of the field-charges in the abstract IFCS field is independent of the given physical interaction (gravitational, electromagnetic, weak, strong); the same transformation rotates the field-charges of all fundamental interactions between their scalar (potential) and vector (kinetic) states appearing in the respective parts of their Hamiltonian – in an interaction-independent isotopic field-charge spin (IFCS or $\Delta$) field. The gravitational eigenvalue equation does not lose its covariant character by its extension with the IFCS section, rather, this extension guarantees its covariance in the presence of a velocity dependent field and IFC-s.